\documentclass[twocolumn,aps,pra,showpacs,raggedbottom,nobalancelastpage,amssymb,superscriptaddress]{revtex4-1}

\usepackage{graphicx}
\usepackage{amsmath}
\usepackage{amssymb}
\usepackage{bm}
\usepackage[usenames]{color}
\usepackage{amsfonts}
\usepackage{appendix}
\usepackage{dsfont}

\definecolor{JanRed}{rgb}{0.81,0.20,0.15}
\definecolor{Nathanblue}{rgb}{0.,0.24,0.51}
\newcommand{\blue}{\color{Nathanblue}}
\newcommand{\red}{\color{JanRed}}

\def\be{\begin{equation}}
\def\ee{\end{equation}}

\def\bs#1{\boldsymbol{#1}}

\begin{document}

\title{{\blue Topological Quantum Matter with Ultracold Gases in Optical Lattices{\red$^{\star}$}}}

\author{N. Goldman}
\email[]{ngoldman@ulb.ac.be\flushleft{\vspace{-0.1cm}\red $\, ^{\star}\,$}Contribution to the Focus Issue on Topological Matter: \\ Nature Physics {\bf 12}, 639-645 (2016). {\blue doi:10.1038/nphys3803}}
\affiliation{CENOLI, Facult{\'e} des Sciences, Universit{\'e} Libre de Bruxelles (U.L.B.), B-1050 Brussels, Belgium}
\author{J. C. Budich}
\affiliation{Institute for Quantum Optics and Quantum Information of the Austrian Academy of Sciences, 6020 Innsbruck, Austria}
\affiliation{Institute for Theoretical Physics, University of Innsbruck, 6020 Innsbruck, Austria}
\author{P. Zoller}
\affiliation{Institute for Quantum Optics and Quantum Information of the Austrian Academy of Sciences, 6020 Innsbruck, Austria}
\affiliation{Institute for Theoretical Physics, University of Innsbruck, 6020 Innsbruck, Austria}


\begin{abstract}Since the discovery of topological insulators, many topological phases have been predicted and realized in a range of different systems, providing both fascinating physics and exciting opportunities for devices. And although new materials are being developed and explored all the time, the prospects for probing exotic topological phases would be greatly enhanced if they could be realized in systems that were easily tuned. The flexibility offered by ultracold atoms could provide such a platform. Here, we review the tools available for creating topological states using ultracold atoms in optical lattices, give an overview of the theoretical and experimental advances and provide an outlook towards realizing strongly correlated topological phases.\end{abstract}

\maketitle


Since the discovery of the  quantum Hall (QH) effect~\cite{PrangeGirvin} and topological insulators~\cite{RMP_TI,RMP_TI2}, intense effort has been devoted to the exploration of novel topological phases of matter. This quest is driven by the prediction of fundamentally new physical phenomena, some of which envisage potential applications~\cite{RMP_TI,RMP_TI2,fractionalChern2,NayakReview}.  Topological phases of matter elude the conventional Landau paradigm of local order parameters associated with spontaneous symmetry breaking.  Instead, topological phases are distinguished by non-local topological invariants, such as Chern numbers: distinct topological phases cannot be smoothly (adiabatically) deformed one into another, even though they may preserve the same symmetries~\cite{RMP_TI,RMP_TI2}.

Among the  physical platforms that have been considered to realize and probe these intriguing phases, \emph{ultracold atoms in optical lattices} appear as promising candidates. These highly controllable and flexible systems consist of dilute gases of neutral atoms, constrained to move in periodic potential patterns created by laser fields~\cite{Bloch:2008gl,Lewenstein,Grimm}.  In these setups, both the band structure~\cite{GrynbergReview} and the strength of the inter-particle interactions~\cite{Chin} can be tuned by  external electromagnetic fields, offering a unique platform  for the realization of unconventional phases of matter~\cite{Bloch:2008gl,Lewenstein,Cooper_review,Dalibard2011,Goldman:2014bv}, including topological phases. Moreover, the high degree of control over the internal atomic structure enables  flexible manipulation and imaging of single atoms~\cite{GreinerMicroscope,Endres,GreinerEntanglement}, thus providing the technology to probe signatures of topology, which may go beyond conventional solid-state experiments.

It is the aim of this Progress Article to review the main tools for the implementation of topological states and band structures with ultracold atoms in optical lattices. We discuss how geometrical and topological properties can manifest in these systems, and describe the measurement techniques that are developed to probe them. We then discuss recent theoretical developments, setting the focus on a dissipative approach that addresses the issue of state preparation. As an outlook, we review promising directions towards reaching  topological superfluids and strongly correlated topological phases in ultracold gases.

\subsection*{The cold-atom toolbox}

We start with a very brief overview of the toolbox that has been developed to create and probe synthetic matter with cold atoms in optical lattices.

The main interface to control atoms through light-matter interaction is the optical  dipole potential~\cite{Grimm,GrynbergReview}, which can be generated by subjecting atoms to laser fields.  It can be expressed as $V (\bs x)\!=\!\alpha  \vert \bs{E} (\bs x) \vert^2$, where $\bs{E} (\bs x)$ denotes the electric field associated with the lasers, $\alpha $ is the polarizability, which typically depends on the laser frequency, and $\bs x$ denotes the position vector. By interfering several beams, rich spatial patterns of light forming adjustable potential-landscapes for atoms can be created. Of great interest are those configurations leading to space-periodic traps, called \emph{optical lattices}, which can form arbitrary geometries (square, honeycomb, \dots) of various  dimensions~\cite{Grimm,Lewenstein,Bloch:2008gl,GrynbergReview}.  These synthetic lattices can be made static, e.g.~using standing waves $E(x)\!\sim\! \cos(q x)$. In the case of a deep lattice, the  dynamics of the atoms is well captured by the  Hubbard Hamiltonian~\cite{Bloch:2008gl,Lewenstein}, a familiar model for a single electronic band,
\be
 \hat H= - J \sum_{\langle m,n \rangle} \hat a^{\dagger}_m \hat a_n + \hat U_{\text{int}},\label{hubbard}
\ee
where $\hat a^{\dagger}_m$ creates an atom at lattice site $m$, $J$ denotes  the constant tunneling matrix element between nearest-neighboring sites $\langle m,n \rangle$, and the interaction term $\hat U_{\text{int}}$ describes on-site (contact) interactions.  Such lattice Hamiltonians can be equally realized for Fermi and Bose gases.

Optical lattices can also be made dynamic. For instance, ``moving" lattices can be obtained by interfering two laser beams with slightly different  frequencies \cite{Stamper1999}. Optical lattices can be rotated~\cite{Cooper_review}, and even \emph{shaken}, e.g.~using piezo-electric actuators~\cite{jotzu2014}. Time-dependent optical lattices constitute a powerful tool for engineering atomic gases with topological properties~\cite{Aidelsburger:2015,jotzu2014}; this ``Floquet-engineering" approach~\cite{Sorenson,Kitagawa,Lindner,Hauke2012,GoldmanDalibard,Zhai,Bukov,Bermudez,Kolovsky,Eckardt2,Cayssol} will be presented below. 

Another important tool is the coherent coupling between different atomic internal states, using laser fields whose frequencies are resonant with specific atomic transitions.
Driving controlled transitions between  internal states can be exploited to manipulate single atoms~\cite{Endres,GreinerEntanglement,Weitenberg}, but also, to  generate artificial gauge potentials~\cite{Juz:2004,Jaksch,Juz:2005,Osterloh:2005} and so-called \emph{synthetic dimensions}~\cite{Celi:2014}, as will be explained below (see \cite{Dalibard2011,Goldman:2014bv} for detailed discussions).

Cold atoms can be visualized by imaging the atomic cloud \emph{in-situ}~\cite{Bloch:2008gl,Endres}. Momentum distributions and band populations can also be obtained through time-of-flight imaging and band-mapping~\cite{Bloch:2008gl}, and the dispersion relation of excitations can be extracted using light scattering~\cite{Stamper1999}. Recently, correlations~\cite{Endres} and entanglement entropy~\cite{GreinerEntanglement} have also been evaluated using single-site resolved images.

 \subsection*{Probing geometry and topology with cold atoms}

We now briefly review the tools that have been developed to reveal geometrical and topological properties of band structures in the cold-atom context. We focus our discussion on 2D atomic systems exhibiting (integer) QH physics, highlighting those experimental probes that are specific to cold gases, and then mention extensions to other topological atomic states. 

The geometrical structure of Bloch bands is captured by the Berry curvature~\cite{Berryoriginal,KarplusLuttinger,MeadReview,Xiao2010}
\be
\Omega= i \left ( \langle \partial_{k_x} u_{\lambda} \vert \partial_{k_y} u_{\lambda} \rangle - \langle \partial_{k_y} u_{\lambda} \vert \partial_{k_x} u_{\lambda} \rangle   \right ),
\ee
where $\vert u_{\lambda} (\bs k) \rangle$ is the Bloch state of the band $\lambda$ at  quasi-momentum $\bs{k}$.  In cold gases, various physical signatures of the Berry curvature can be probed:~For instance, the anomalous (transverse) velocity of a wave-packet under the action of a force~\cite{PriceCooper}, or the Aharonov-Bohm phase acquired by a wave-packet performing a loop in $\bs k$-space (reflecting that the Berry curvature acts as a ``magnetic field" in $\bs k$-space)~\cite{Duca}. For two-band systems, the Berry curvature can be simply expressed in terms of the momentum distribution, and hence, it can be directly reconstructed from time-of-flight images~\cite{Alba,HaukeChern}. These different probing strategies have been successfully implemented in recent experiments, see Refs.~\cite{jotzu2014,Aidelsburger:2015,Duca,Flaschner}.

\begin{figure}
\includegraphics[width=8.5cm]{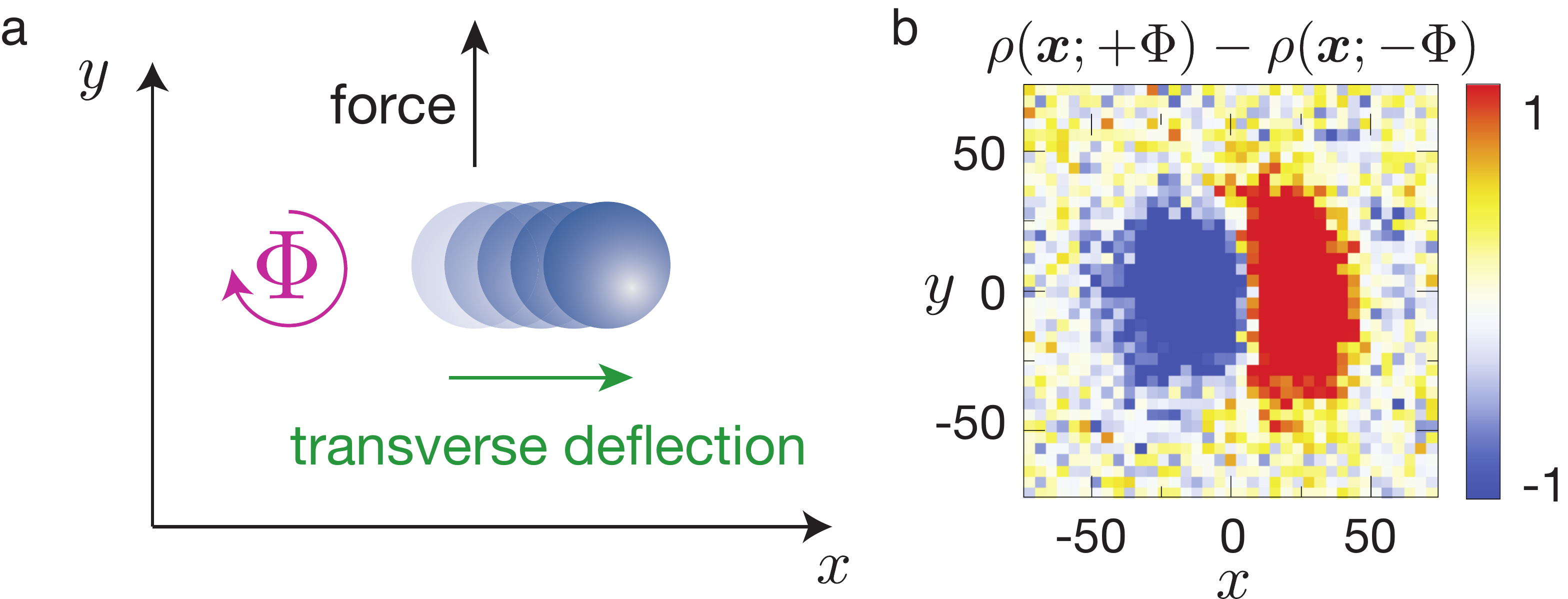}
\vspace{-0.cm} \caption{{\bf Measurement of the Chern number.} (a)  Loading an atomic cloud into Bloch bands with non-zero Chern numbers (e.g.~by creating a synthetic uniform magnetic flux $\Phi$ in an optical lattice~\cite{Dalibard2011,Goldman:2014bv}), and subjecting it to a constant force, produces a transverse drift, in direct analogy with the electrical Hall effect. When atoms are uniformly loaded into a single Bloch band with Chern number $\mathcal C$, the center-of-mass velocity along the transverse direction is directly proportional to the Chern number $\mathcal C$; see text and~\cite{Dauphin:2013,Price:2016}. (b) In the experiment realized in Munich~\cite{Aidelsburger:2015}, the transverse drift of the cloud was imaged \emph{in-situ} in response to an applied force $\bs{F}=F_y \bs{1}_y$. Here we show a typical image of the particle density $\rho (\bs x)$, which was produced by subtracting the data obtained for opposite values of the synthetic flux $\pm \Phi$. The measured transverse drift allowed one to extract an approximate value for the Chern number of the populated band, $\mathcal C\!=\!0.99(5)$, in agreement with the theoretical prediction $\mathcal C\!=\!1$.}\label{Fig_one}\end{figure}

Topological invariants are global properties and can often be expressed as integrals over local geometric quantities. For example, the genus (i.e.~number of handles) of a closed two-dimensional (2D) surface is determined by integrating its Gaussian curvature~\cite{Nakahara}. Similarly, the topology of a Bloch band in 2D can be characterized by the integrated Berry curvature over the entire Brillouin zone (BZ),
\be
\mathcal C= \frac{1}{ 2\pi} \int_{\text{BZ}} \Omega \, \, \text{d}^2 k \quad \in \mathbb{Z}.
\ee
This so-called \emph{Chern number} \cite{TKNN1982} of the band is at the heart of the integer  QH effect in electronic systems: the topologically quantized Hall conductance associated with a completely filled band is given by $\sigma_H\!=\!\sigma_0 \mathcal C$, where $\sigma_0\!=\!e^2/h$ is the quantum of conductance~\cite{PrangeGirvin,TKNN1982}. When atoms are uniformly loaded into a Bloch band with Chern number $\mathcal C$, and subsequently subjected to a force of strength $F$, the center-of-mass velocity along the transverse direction is given by $v_{\text{c.m.}}\!=\! \mathcal C  A_{\text{cell}} F  / h$, where $A_{\text{cell}}$ denotes the unit-cell area, and $h$ is Planck's constant~\cite{Dauphin:2013,Price:2016}, see  Fig.~\ref{Fig_one}(a). This quantized center-of-mass (COM) response is an unambiguous manifestation of $\mathcal C$ in the \emph{bulk}, as opposed to \emph{edge} currents detected in Hall bars. This drift can be directly observed \emph{in-situ}~\cite{Aidelsburger:2015}, see  Fig.~\ref{Fig_one}(b). As discussed in Ref.~\cite{Price:2016}, such COM observables could even detect quantized electromagnetic responses not captured by conductivity measurements. Furthermore, the Chern number $\mathcal C$ can also be observed through the aforementioned Berry-curvature-reconstruction schemes~\cite{PriceCooper,Alba,HaukeChern,Duca,Flaschner}, or by measuring the spin polarization of an atomic cloud at highly-symmetric points of the Brillouin zone~\cite{Liu:2013,2DSOC2:2015}. Finally, we point out that a many-body Chern number, as defined in interacting systems, may be probed by extending the interferometry scheme of Ref.~\cite{Duca} to mobile impurities bound to quasiparticles~\cite{Chernimpurity}. 

The bulk-edge correspondence guarantees the existence of chiral edge modes whenever a band structure displays Bloch bands with non-zero Chern numbers~\cite{Halperin1982, RMP_TI,RMP_TI2}. In cold gases, the high-resolution addressing techniques offer the possibility of directly loading atoms into these edge states and to visualize their time-evolution in real space~\cite{GoldmanPNAS,MuellerEdge}; see Refs.~\cite{Atala:2014,Mancini:2015,Stuhl:2015} for the experimental detection of edge motion. Moreover, a complete reconstruction of the edge-modes dispersions could be performed using Bragg spectroscopy~\cite{Liuspectro,Stanescu,Goldmanspectro}. We note that these techniques may also be adapted to investigate the edge modes of strongly-correlated states.

The schemes described above can be simply generalized to probe the physics of time-reversal-invariant topological states~\cite{RMP_TI,RMP_TI2}, both in 2D and 3D, so as to reveal the corresponding $Z_2$ invariants and detect helical edge modes. Besides, unambiguous signatures of axion-electrodynamics~\cite{Bermudez2010} -- a genuine property associated with 3D topological insulators~\cite{RMP_TI,RMP_TI2} -- and the detection of  Weyl fermions~\cite{WeylAtoms} have been proposed in the cold-atom context. Finally, certain aspects of higher-dimensional topological states, such as 4D  QH responses, are accessible in cold gases~\cite{4Datoms:2015,Price:2016}, through the concept of synthetic dimensions \cite{Celi:2014} (see discussion below). 

A series of proposals for detecting (possibly strongly-correlated) topological states are discussed in the reviews~\cite{Cooper_review,Goldman:2014bv}.

 \subsection*{Artificial gauge fields for cold atoms \\ in optical lattices}

\begin{figure}
\includegraphics[width=8.3cm]{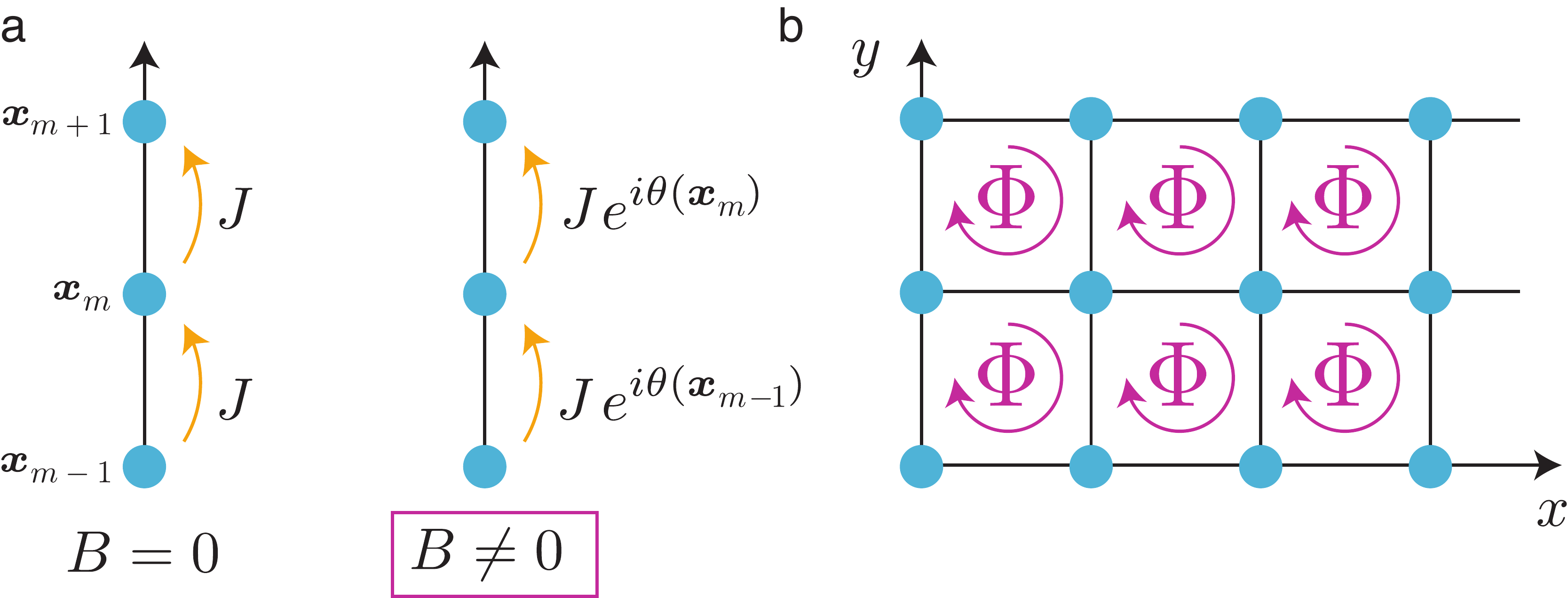}
\vspace{-0.cm} \caption{{\bf  The Peierls substitution and the Aharonov-Bohm effect.} (a) In the absence of  external field ($B\!=\!0$), the motion of a particle on a lattice  can be characterized by  a constant real-valued tunneling parameter $J$, see Eq.~\eqref{hubbard}. When applying an external magnetic field ($B\!\ne\!0$), the Peierls substitution modifies the tunneling matrix elements, through the complex phase-factors $\exp [i \theta (\bs x)]$ defined in Eq.~\eqref{Peierls}. (b)  On a 2D lattice, the wave function of a particle performing a loop around a plaquette may acquire a non-zero Aharonov-Bohm phase-factor $\exp (i \Phi)$, where $\Phi\!=\! \oint \bs{A} \cdot \text{d} \bs{l}$ is the magnetic flux through the plaquette; the latter is obtained by summing the Peierls phases $\theta (\bs x)$ in Eq.~\eqref{Peierls} around this plaquette. A 2D lattice penetrated by a uniform flux per plaquette $\Phi$ corresponds to the emblematic Harper-Hofstadter model~\cite{Hofstadter}, and it exhibits Bloch bands with non-zero Chern numbers~\cite{RMP_TI,RMP_TI2}.}\label{Fig_zero}\end{figure}

In solid-state systems,  prominent mechanisms inducing topological Bloch bands  include spin-orbit coupling and externally applied magnetic fields~\cite{RMP_TI,RMP_TI2}. Formally, these \emph{gauge fields} affect the tunneling of electrons within the crystal through the Peierls substitution~\cite{Luttinger,Hofstadter,Xiao2010}.  For instance,  an external magnetic field $\bs B \!=\! \bs{\nabla}\!\times\!\bs{A}$ modifies the Hubbard Hamiltonian in Eq.~\eqref{hubbard} through the Peierls substitution (hereafter we set $\hbar\!=\!e\!=\!1$ unless otherwise stated)
\be
 - J \sum_{\langle m,n \rangle} \hat a^{\dagger}_m \hat a_n \!\longrightarrow\! -J\sum_{\langle m,n \rangle}  \hat a^{\dagger}_m \left ( e^{i \int_{\bs{x}_n}^{\bs x_m}\bs{A}\cdot \text{d} \bs{x}} \right ) \hat a_n , \label{Peierls}
\ee 
where $J$ denotes the tunneling matrix element between lattice sites $\bs{x}_{m}$ and  $\bs{x}_{n}$ in the absence of a field, and where $\bs{A}$ is the electromagnetic gauge potential, see Fig.~\ref{Fig_zero}(a).  The ``Peierls phase-factor" in Eq.~\eqref{Peierls} directly determines the Aharonov-Bohm phase acquired by a particle encircling a unit cell of the lattice [Fig.~\ref{Fig_zero}(b)]. The Peierls substitution in Eq.~\eqref{Peierls} can readily be generalized to the case of non-Abelian gauge fields~\cite{Xiao2010}, e.g.~spin-orbit coupling, by means of a path-ordered integral. 

 In gases of neutral atoms, analogs of the gauge field $\bs A$ appearing in \eqref{Peierls} can be generated artificially, using the atom-light interaction~\cite{Juz:2004,Jaksch,Juz:2005,Osterloh:2005,Dalibard2011,Goldman:2014bv}. In the following paragraphs, we detail various ways of engineering such \emph{artificial gauge fields} that allow for the realization of topological matter with cold atoms.

 The schemes presented below assume that cold atoms are trapped in deep optical lattices. In this \emph{tight-binding} regime [Eqs.~\eqref{hubbard} and \eqref{Peierls}], artificial magnetic fields appear as effective fluxes defined at each plaquette [Fig.~\ref{Fig_zero} (b)]. We emphasize that there exists a complementary approach, based on ``optical flux lattices"~\cite{CooperFluxLattice}, which realizes effective magnetic flux densities that are continuous functions of position. We note that artificial magnetic fields and spin-orbit couplings can also be produced in the absence of a lattice, e.g. using rotation~\cite{Cooper_review} or atom-light coupling~\cite{Dalibard2011,Goldman:2014bv}.

\subsection*{Laser-induced tunneling}

 Historically, laser-induced tunneling~\cite{Jaksch1998,Ruostekoski} was the first method proposed to generate artificial magnetic fields in optical lattices~\cite{Jaksch,Gerbier}. Consider a gas of atoms trapped in an optical lattice, and let us assume that an atom in an internal state $\vert 1 \rangle$ is present at a lattice site $\bs x_1$, while an atom in state $\vert 2 \rangle$ is present at a neighboring site $\bs x_2$. \emph{Laser-induced tunneling} can then be activated between the two sites, by coupling the two internal states $\vert 1 \rangle\!\leftrightarrow\!\vert 2 \rangle$ with a resonant laser field~\cite{Jaksch1998,Ruostekoski,Jaksch,Gerbier}. The resulting effective tunneling matrix element $J_{\text{eff}}\!=\!\vert J_{\text{eff}} \vert \exp [i \bs{q} \cdot \bs x]$ includes a complex phase-factor, which is directly related to the wave vector $\bs q$ of the coupling field~\cite{Jaksch}. Hence, the effective tunneling $J_{\text{eff}}$ takes the form of the aforementioned Peierls substitution [Eq.~\eqref{Peierls}]: atoms  behave as if they were (charged) particles subjected to gauge fields~\cite{Dalibard2011,Goldman:2014bv}. This method was proposed to simulate synthetic magnetic fields~\cite{Jaksch,Gerbier} and spin-orbit coupling~\cite{Goldman2010,Bermudez2010,Liu2014} in optical lattices, in view of creating topological insulators and superfluids in atomic gases. It was implemented very recently in an experiment realizing artificial spin-orbit coupling in a 2D optical lattice~\cite{2DSOC2:2015}; with the addition of a controllable Zeeman field, this configuration led to an anomalous-QH system with Chern bands~\cite{RMP_TI}, whose topology was directly revealed by measuring spin polarization~\cite{2DSOC2:2015} at highly symmetric points of the Brillouin zone~\cite{Liu:2013}.

Complementary to the engineering of Chern bands for itinerant atoms, coupling internal atomic states with lasers can also be employed to engineer interactions between localized spins in magnetic gases, resulting in Chern bands for hardcore bosons \cite{YaoFlatDipolar}. Besides, the orbital degrees of freedom can also be exploited in optical lattices, where Chern bands can emerge through interactions~\cite{Sun2011,DauphinMott}.

\begin{figure*}
\includegraphics[width=17.5cm]{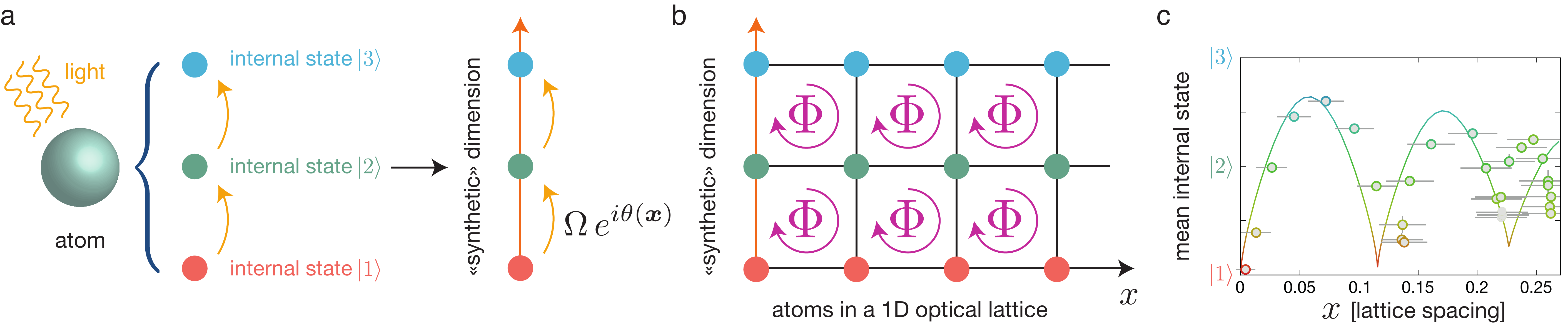}
\vspace{-0.cm} \caption{{\bf  Exploiting a synthetic dimension.} (a) The synthetic dimension approach consists in interpreting a set of atomic internal states as fictitious lattice sites, aligned along an extra spatial dimension. Driving transitions between the internal states allows for ``hopping" along the synthetic dimension, which is then characterized by the tunneling matrix element $\kappa\!=\!\Omega\exp [i \theta (\bs x)]$, where the Peierls phase-factor is directly related to the wave-vector of the driving laser field~\cite{Celi:2014}. (b) Loading atoms into a 1D optical lattice, and driving transitions between the internal states, can be used to simulate lattice systems of two spatial dimensions. By adjusting the Peierls phase-factors $\exp [i \theta (\bs x)]$, synthetic magnetic fluxes $\Phi$ can be created within this fictitious 2D lattice. (c) Experimental observation of edge skipping orbits, in a synthetic 3-leg ladder (using three internal states of $^{173}$Yb atoms) subjected to a uniform synthetic flux~\cite{Mancini:2015}.}\label{Fig_two}\end{figure*}

\subsection*{The synthetic dimensions approach}

The concept of synthetic dimensions~\cite{Celi:2014} offers a novel playground for the experimental exploration of gauge fields and topological states in cold gases. It consists in interpreting a set of internal states of an atom, e.g.~Zeeman sublevels of a hyperfine state~\cite{Mancini:2015,Stuhl:2015}, as fictitious lattice sites; this defines an extra ``spatial" dimension coined \emph{synthetic dimension}, see Fig.~\ref{Fig_two} (a). In this picture, driving transitions between different internal states, e.g.~using resonant laser fields, corresponds to inducing ``hopping" processes along the synthetic dimension. Interestingly, loading atoms into a (real) $N$-dimensional optical lattice then potentially allows one to simulate systems of $(N\!+\!1)$ spatial dimensions. Hence, this technique offers a versatile tool to explore quantum effects associated with higher-dimensions, such as the 4D QH effect~\cite{4Datoms:2015}.

Specifically, consider the laser-coupling between two internal states of an atom, $\vert 1 \rangle\!\leftrightarrow\!\vert 2 \rangle$; the corresponding coupling matrix element is of the form $\kappa \!=\!\Omega\exp [i \bs{q} \cdot \bs x]$, where $\Omega$ denotes the coupling strength and $\bs q$ denotes the wave-vector of the coupling field~\cite{Dalibard2011,Goldman:2014bv}. In the synthetic dimension picture, the quantity $\kappa$ represents the tunneling matrix element between the fictitious sites $``1"$ and $``2"$; see Fig.~\ref{Fig_two} (a). Interestingly, and similarly to the laser-induced-tunneling scheme discussed above, the fictitious tunneling $\kappa$ contains a complex phase-factor, which can then be exploited to simulate artificial gauge fields in a simple and practical manner~\cite{Celi:2014,4Datoms:2015}, see Fig.~\ref{Fig_two} (b).

Synthetic dimensions were recently investigated in two independent experiments~\cite{Mancini:2015,Stuhl:2015}. Atoms were loaded into a 1D optical lattice, while a laser-coupling scheme was added to drive coherent transitions between three internal atomic states: this setup effectively reproduced a 3-leg ladder, in the synthetic-dimension picture, see Fig.~\ref{Fig_two}~(b). By adjusting the wave-vector of the coupling field $\bs q$, the experimentalists generated an artificial flux threading the ladder. The corresponding band structure exhibits chiral edge modes~\cite{Celi:2014}, reminiscent of the edge states in the QH effect~\cite{Halperin1982}. These edge modes are characterized by semi-classical skipping orbits at the edges of the synthetic ladder, which can be directly imaged through state-resolved images of the cloud, as was experimentally demonstrated in Refs.~\cite{Mancini:2015,Stuhl:2015}, see Fig.~\ref{Fig_two} (c). 

In order to engineer topological band structures using the synthetic-dimension approach, atomic species with many addressable internal states (e.g.~Yb, Sr) are required. This is because a proper bulk region within the artificial dimension is crucial to limit undesired finite-size effects (e.g.~the overlap of chiral edge modes associated with opposite edges). An additional coupling to connect the extremal internal states may be used to apply periodic boundary conditions in the synthetic dimension \cite{Celi:2014}. Finally, we point out two qualitative differences between ordinary crystalline systems and systems involving a synthetic dimension. First, in the latter case, the hopping parameters are typically inhomogeneous due to the Clebsch-Gordan coefficients associated with the atomic transitions. Second, the interactions are generically infinite-ranged (i.e.~spin-dependent interactions are negligible). While the resulting extended-Hubbard model displays interesting phases~\cite{Barbarino}, it is still debated whether such systems could host fractional QH states~\cite{Lacki}.

\subsection*{Shaking atoms into topological matter}
Complementary to the schemes discussed so far, which rely on the possibility of addressing the internal structure of atoms with light, there exists an even more general strategy to engineer topological band structures in quantum systems, which is commonly called \emph{Floquet engineering}~\cite{Sorenson,Kitagawa,Lindner,Hauke2012,GoldmanDalibard,Zhai,Bukov,Bermudez,Kolovsky,Eckardt2,Cayssol}. This approach, which is based on applying time-periodic modulations to quantum systems, can be summarized as follows. Consider a static system described by a Hamiltonian $\hat H_0$ that is driven by a time-periodic modulation $\hat V (t)$, whose period $T\!=\!2 \pi/\omega$ is assumed to be much smaller compared to any characteristic time scale in the problem. In this high-frequency regime, the dynamics is generally well captured by an effective Hamiltonian $\hat H_{\text{eff}}$, which stems from a rich interplay between the static and time-dependent parts of the total Hamiltonian $\hat H_0\!+\!\hat V (t)$, see Refs.~\cite{GoldmanDalibard,Bukov}. In this way, a target Hamiltonian (e.g.~one that leads to a topological band structure) can be engineered through an appropriate choice of the static system and driving protocol. As we explain below, this approach was recently implemented in cold-atom  setups, where time-dependent optical lattices were used to engineer Bloch bands with non-zero Chern numbers~\cite{jotzu2014,Aidelsburger:2015}.

\begin{figure}
\includegraphics[width=8.5cm]{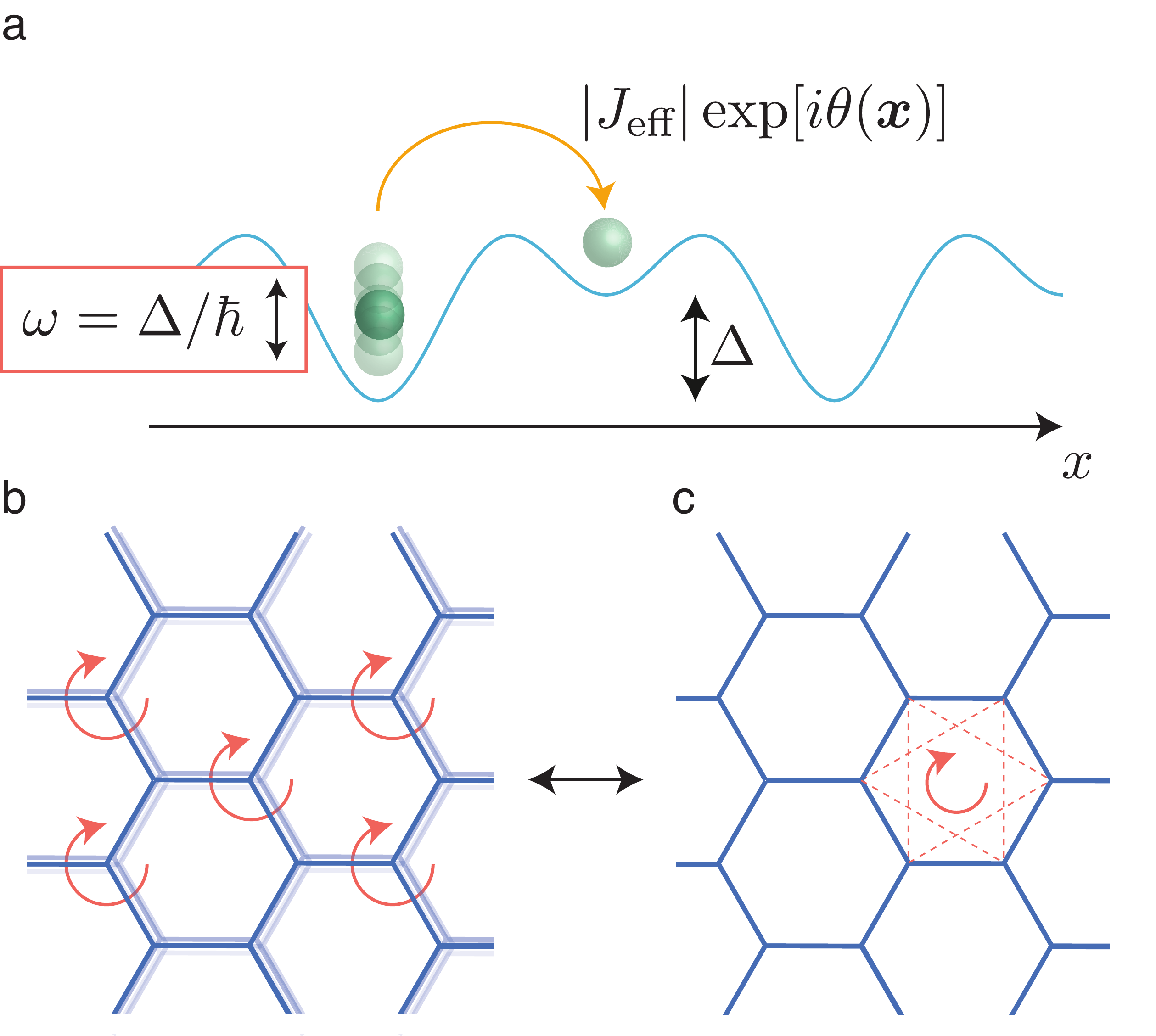}
\vspace{-0.cm} \caption{{\bf Floquet engineering with cold atoms.} (a) Tunneling can be induced in optical superlattices, by modulating the potential resonantly with respect to the energy offsets $\Delta$ between neighboring sites. Considering a modulation potential of the form $V(\bs x, t)\!=\!V_0 \cos [\omega t + \theta (\bs x)]$, the resulting tunneling matrix elements are  $J_{\text{eff}}\!=\!\vert J_{\text{eff}} \vert \exp [i \theta(\bs x)]$. This method allows one to induce Peierls phase-factors, and hence, artificial magnetic fluxes in 2D optical lattices~\cite{aidelsburger2013,miyake2013,Aidelsburger:2015,Kennedy}. (b) Shaking a honeycomb optical lattice circularly is formally equivalent to subjecting graphene to circularly-polarized light: this configuration reproduces the so-called Haldane model~\cite{Haldane}, shown in (c), where effective next-nearest-neighbor hopping matrix elements are associated with a well-defined chirality (stemming from the shaking~\cite{jotzu2014}). }\label{Fig_one_bis}\end{figure}

A series of experiments~\cite{aidelsburger2013,miyake2013,Aidelsburger:2015,Kennedy}  developed a scheme based on resonant modulations~\cite{Bermudez,Kolovsky}. It starts with a static optical superlattice (or a Wannier-Stark ladder) characterized by energy offsets $\Delta$ between neighboring sites, large enough to inhibit bare tunneling $J\!\ll\! \Delta$. A time-modulated optical potential $\hat V (t)$ is then added, with resonant frequency $\omega\!=\!\Delta/\hbar$, to restore the tunneling in a controlled way, see Fig.~\ref{Fig_one_bis} (a). In direct analogy with the laser-induced-tunneling scheme described above, the effective tunneling matrix elements then acquire Peierls phase-factors, $J_{\text{eff}}\!=\!\vert J_{\text{eff}} \vert \exp [i \theta(\bs x)]$, which are directly related to the phase $\theta(\bs x)$ of the modulated potential~\cite{Bermudez,Kolovsky}. In this way, the moving potential $\hat V (t)$ can be designed so as to engineer synthetic gauge fields in 2D optical lattices~\cite{Kolovsky}. The experiments reported in Refs.~\cite{aidelsburger2013,miyake2013,Aidelsburger:2015,Kennedy} realized strong synthetic magnetic fluxes and studied their effects on atomic gases. In Refs.~\cite{aidelsburger2013,Aidelsburger:2015}, a constant flux per plaquette $\Phi\!=\!\pi/2$ was engineered throughout the lattice, leading to Bloch bands with non-zero Chern numbers. This allowed for the first Chern-number measurement with cold gases~\cite{Aidelsburger:2015}, which was achieved by loading bosonic atoms into a Bloch band with Chern number $\mathcal C\!=\!1$ and subjecting them to a constant force; the Chern number was then extracted by measuring the transverse (Hall) drift of the cloud~\cite{Dauphin:2013,Price:2016}, see Figure~\ref{Fig_one}.  The experiment~\cite{Aidelsburger:2015} also revealed a  significant heating, attributed to the periodic driving, through dynamical measurements of  excitations to higher bands. This pointed out the current limitations of this scheme in view of stabilizing topological states. Similar center-of-mass drift measurements were recently performed~\cite{Lohse,Nakajima,Lu2015}  to extract the Zak phase and Chern number of 1D (Thouless) quantum pumps~\cite{Thouless1983,Xiao2010}.

Shaking a 2D optical lattice in an off-resonant manner also constitutes a powerful method by which effective magnetic fluxes and topological band structures can be created~\cite{jotzu2014,Struck}. In particular, in the high-frequency regime, the effective Hamiltonian $\hat H_{\text{eff}}$ of a circularly shaken honeycomb optical lattice~\cite{jotzu2014,Zhai} becomes equivalent to the emblematic Haldane model~\cite{Haldane}, in direct analogy with graphene irradiated by circularly-polarized light~\cite{Cayssol}; see Figure~\ref{Fig_one_bis} (b,c). This allows for the realization of the anomalous Hall effect in cold atoms, as was experimentally demonstrated in Ref.~\cite{jotzu2014} through the observation of an anomalous velocity~\cite{KarplusLuttinger,Xiao2010} in response to an applied force.

\subsection*{Topological superfluids and Majorana bound states}
The occurrence of quasiparticles with non-Abelian statistics, called non-Abelian anyons, has first been predicted in certain fractional QH states \cite{MooreRead}, and later in time-reversal-breaking superconductors with $p+ip$ pairing in 2D \cite{ReadGreen}, as well as $p$-wave pairing in 1D \cite{Kitaev2001}. The specific anyons occurring in these systems are known as Majorana bound states (MBS), due to their algebraic similarities with the real solutions to the relativistic Majorana equation. More recently, it has become clear that $p+ip$ and $p$-wave superconductors can be induced in conventional (proximity induced) superconductors due to the combination of spin-orbit coupling and Zeeman splitting \cite{SauRashbaZeeman}. With the recent advances in the experimental realization of synthetic spin-orbit coupling in ultracold quantum gases \cite{SpielmanReview,2DSOC2:2015,2DSOCNatPhys}, all the individual ingredients for synthetic topological superfluids in fermionic quantum gases are in place and several concrete proposals, both for the 2D $p+ip$ superfluids \cite{pwaveDasSarma,pwaveLewenstein,pwaveMelo} and proximity induced 1D $p$-wave superfluids \cite{Tewaripip,Jiang11,Nascimbene12,Kraus13}, have been put forward. Once experimentally realized, the high degree of experimental control over these systems enables new approaches for the direct observation of MBS via braiding \cite{KrausBraiding}.

\subsection*{Dissipative preparation of topological states}
So far we have discussed several tools to engineer Hamiltonians, the ground states of which have topologically non-trivial properties. A complementary approach, in which desired many-body states are directly targeted, is provided by the concept of dissipative state preparation \cite{DiehlDissPrep,Verstraete2008}. Intuitively dissipation is expected to increase the entropy of a system thus having a detrimental effect on ordering phenomena. However, the flexibility to engineer the interaction of cold atom systems with their environment allows one to think about dissipative processes as a resource to control quantum many-body systems in a non-equilibrium fashion. That way, dissipation can be harnessed to prepare interesting states of quantum matter as steady states of a master equation governing the open quantum system dynamics. For a weak coupling to a Markovian bath, which in many cases  represents a good approximation for atoms coupled to a continuum of radiation modes, the master equation is of Lindblad form and reads as \cite{Lindblad1976}
\begin{align}
\text{d} \hat \rho/\text{d} t  = i\left[\hat \rho, \hat H\right]+\sum_j\left(\hat L_j \hat \rho \hat L_j^\dag -\frac{1}{2}\left\{\hat L_j^\dag \hat L_j, \hat\rho\right\}\right),
\label{eqn:Lindblad}
\end{align}
where $\hat \rho$ denotes the reduced density matrix of the system, and where the incoherently acting Lindblad operators $\hat L_j$ (also called jump operators) account for the system-bath coupling; the dissipative channels are labelled by $j$, and are often directly related to the degrees of freedom of a lattice system \cite{DiehlTopDiss,BardynTopDiss}. Steady states $\hat \rho_s$ are defined by  $\text{d} \hat \rho_s/\text{d} t \!=\!0$. In this dissipative context, the counterpart to an energy gap protecting a Hamiltonian ground state is provided by a damping gap \cite{DiehlTopDiss,BardynTopDiss}, defined as the smallest rate at which deviations from $\hat \rho_s$ are damped out. Early works devoted to dissipative state preparation \cite{DiehlTopDiss} mainly focused on purely dissipative dynamics by assuming the system Hamiltonian $H$ to be zero. Then, pure steady states $\lvert D\rangle\langle D\rvert$, also referred to as dark states, are simply characterized by $\hat L_j \lvert D\rangle\!=\!0$ for all $j$. Hence, if the desired topological state can be represented as the (unique) ground state $\lvert G\rangle$ of the so-called parent Hamiltonian $\hat H_p\!=\!\sum_j \hat A_j^\dag \hat A_j$, realizing $\hat L_j\!=\!\hat A_j$ as jump operators, will make $\lvert G\rangle$ the (unique) dark state. In the context of topological phases, this approach has been pioneered in Ref.~\cite{DiehlTopDiss}, where a scheme for the dissipative preparation of a 1D topological superconductor \cite{Kitaev2001} with a pair of spatially-separated MBS forming a decoherence-free subspace has been proposed.

Quite remarkably, the modification of the bulk-boundary correspondence in open quantum systems \cite{BardynTopDiss} can lead to phenomena that have no direct analog in Hamiltonian systems \cite{Bardyn2012}. For example, unpaired MBS can, in systems with a topologically trivial bulk, form decoherence-free subspaces, the dissipative analog of degenerate ground states~\cite{Bardyn2012}. The concept of topology by dissipation has formally been extended to higher spatial dimensions and various symmetry classes in Ref.~\cite{BardynTopDiss} for Gaussian, fermionic models. However, there is a fundamental competition between topology and locality \cite{chernalgebraic,DissCI} representing a major caveat for the dissipative preparation of chiral topological phases such as Chern insulators: No exponentially localized set of Lindblad operators $\hat L_j$ can be found that leads to a dark state $\lvert D\rangle$ with a non-vanishing Chern number. This issue has been addressed in Ref.~\cite{DissCI}, where a generic mechanism to prepare a mixed topological state \cite{BardynTopDiss,DissCI,TopDens} corresponding to a Chern insulator at finite temperature has been proposed, based on a local system-bath coupling. In this framework, the topology of the mixed steady state is determined by qualitative features of the system-bath interaction, while going towards a pure steady state, the counterpart of reaching zero temperature in a Hamiltonian system, requires some fine-tuning \cite{DissCI}.

\subsection*{Towards strongly correlated topological phases}
In addition to the possibility of engineering single particle Hamiltonians, the atomic physics toolbox naturally provides us with the means to flexibly tune complex many-body interactions~\cite{Lewenstein,Bloch:2008gl,Chin}. The paramount goal of such quantum simulators is the physical realization and control of quantum many-body systems that cannot be efficiently simulated on a classical computer. In the context of topological states of matter, strongly correlated phases such as fractional quantum Hall (FQH) states \cite{PrangeGirvin} and spin liquids \cite{Wen2007} represent intriguing candidates. Their experimental realization in synthetic material systems could provide new physical insights that are hard to access, both in conventional materials and in numerical simulations of small-size model systems. A primary example along these lines is offered by the possibility to directly observe characteristic entanglement signatures in ultracold atomic gases \cite{HannesEntanglementGrowth}, as has very recently been experimentally demonstrated with quantum gas microscope techniques \cite{GreinerEntanglement}.

FQH physics has been studied for many years, both theoretically and experimentally, in strongly correlated 2D electron gases subjected to a strong perpendicular magnetic field~\cite{PrangeGirvin}. More recently, intense interest has been  aroused by the possibility of realizing FQH states in lattice systems: the fractional Chern insulators (see Ref. \cite{fractionalChern2} for a recent review). The basic ingredients for lattice FQH states are almost {\emph{flat}}  (dispersionless) energetically isolated bands with a {\emph{non-vanishing Chern number}} that are partially filled with {\emph{interacting}} particles. Most interestingly, bands with higher Chern number $\mathcal C>1$ can give rise to FQH states that have no natural analog in conventional FQH systems based on continuous Landau levels. Recently, numerous proposals~\cite{PalmerJaksch,SorensenHafezi,MollerCooper,KapitMueller,NielsenCirac,YaoFlatDipolar,CooperFluxFlat} to realize FQH states in Chern bands with ultracold atoms, e.g. using optical flux lattices \cite{CooperFluxFlat} and dipolar spin systems \cite{YaoFlatDipolar} have been reported. The tunability of interactions in such systems opens up the possibility to realize various new FQH states, as indicated by numerical simulations (see e.g. Ref.~\cite{MollerHigherChern}). 

Very recently, Ref.~\cite{Dai} reported on the implementation of a minimal toric-code Hamiltonian with cold atoms; this setup exhibits  fractional (anyonic) statistics, an unambiguous signature of topological phases~\cite{PrangeGirvin,Kitanew}.

Towards the implementation of spin liquids, tunable dipole-dipole interactions in Rydberg atoms have recently been employed to develop a flexible toolbox for the synthetic realization of frustrated quantum magnetism \cite{AlexFQM,PohlFQM}. In particular, an experimentally feasible scheme for the realization of quantum spin ice has been reported \cite{AlexQSI}.

Remarkable progress has been made regarding the quantum engineering of many-body Hamiltonians, which could potentially lead to strongly-correlated topological phases~\cite{Dalibard2011,Goldman:2014bv,Lewenstein,Bloch:2008gl,Cooper_review}. However, a major challenge is still the preparation of states with sufficiently low temperature or, more generally, states with sufficiently low entropy, in view of making exotic features (e.g.~fractionalized excitations and topological entanglement entropies) experimentally accessible. The notion of dissipative state preparation via system bath engineering discussed above represents a possible direction to overcome these issues.

{\emph{Acknowledgment --}} The authors would like to acknowledge M. Aidelsburger, I. Bloch, L. Fallani, and M. Mancini for providing experimental data. N.G. is financed by the FRS-FNRS Belgium and by
the BSPO under PAI Project No. P7/18 DYGEST. This work has also been supported
by the ERC synergy grant UQUAM.


\begin{thebibliography}{39}
\expandafter\ifx\csname natexlab\endcsname\relax\def\natexlab#1{#1}\fi
\expandafter\ifx\csname bibnamefont\endcsname\relax
  \def\bibnamefont#1{#1}\fi
\expandafter\ifx\csname bibfnamefont\endcsname\relax
  \def\bibfnamefont#1{#1}\fi
\expandafter\ifx\csname citenamefont\endcsname\relax
  \def\citenamefont#1{#1}\fi
\expandafter\ifx\csname url\endcsname\relax
  \def\url#1{\texttt{#1}}\fi
\expandafter\ifx\csname urlprefix\endcsname\relax\def\urlprefix{URL }\fi
\providecommand{\bibinfo}[2]{#2}
\providecommand{\eprint}[2][]{\url{#2}}

\bibitem[{\citenamefont{Prange and Girvin}(1990)}]{PrangeGirvin}
\bibinfo{author}{\bibfnamefont{R.}~\bibnamefont{Prange}} \bibnamefont{and}
  \bibinfo{author}{\bibfnamefont{S.}~\bibnamefont{Girvin}},
  \emph{\bibinfo{title}{The Quantum Hall Effect}}
  (\bibinfo{publisher}{Springer}, \bibinfo{year}{1990}).




\bibitem[{\citenamefont{Hasan and Kane}(2010)}]{RMP_TI}
\bibinfo{author}{\bibfnamefont{M.~Z.} \bibnamefont{Hasan}} \bibnamefont{and}
  \bibinfo{author}{\bibfnamefont{C.~L.} \bibnamefont{Kane}}, \bibinfo{title}{Colloquium:~Topological insulators},
  \bibinfo{journal}{Rev. Mod. Phys.} \textbf{\bibinfo{volume}{82}},
  \bibinfo{pages}{3045} (\bibinfo{year}{2010}).

\bibitem[{\citenamefont{Qi and Zhang}(2011)}]{RMP_TI2}
\bibinfo{author}{\bibfnamefont{X.-L.} \bibnamefont{Qi}} \bibnamefont{and}
  \bibinfo{author}{\bibfnamefont{S.-C.} \bibnamefont{Zhang}}, \bibinfo{title}{Topological insulators and superconductors},
  \bibinfo{journal}{Rev. Mod. Phys.} \textbf{\bibinfo{volume}{83}},
  \bibinfo{pages}{1057} (\bibinfo{year}{2011}).



\bibitem{fractionalChern2} E. J. Bergholtz and Z. Liu, Topological flat band models and fractional Chern insulators, International Journal of Modern Physics B, {\bf 27}(24), 1330017 (2013).

\bibitem{NayakReview} C. Nayak, S. H. Simon, A. Stern, M. Freedman, and S. Das Sarma, Non-Abelian anyons and topological quantum computation, Rev. Mod. Phys. {\bf 80}, 1083 (2008).


\bibitem[{\citenamefont{Bloch et~al.}(2008)\citenamefont{Bloch, Dalibard, and
  Zwerger}}]{Bloch:2008gl}
\bibinfo{author}{\bibfnamefont{I.}~\bibnamefont{Bloch}},
  \bibinfo{author}{\bibfnamefont{J.}~\bibnamefont{Dalibard}}, \bibnamefont{and}
  \bibinfo{author}{\bibfnamefont{W.}~\bibnamefont{Zwerger}}, \bibinfo{title}{Many-body physics with ultracold gases},
  \bibinfo{journal}{Rev. Mod. Phys.} \textbf{\bibinfo{volume}{80}},
  \bibinfo{pages}{885} (\bibinfo{year}{2008}).

\bibitem{Lewenstein} M. Lewenstein, A. Sanpera and V. Ahufinger, \emph{Ultracold Atoms in Optical Lattices: Simulating quantum many-body systems}, Oxford University Press; 1 edition (May 4, 2012).

\bibitem{Grimm} R. Grimm, M. Weidem\"uller, and Y. B. Ovchinnikov, Optical Dipole Traps for Neutral Atoms, Adv. At. Mol. Opt. Phys. {\bf 42}, 95-170 (2000).

\bibitem{GrynbergReview}  G. Grynberg and C. Robilliard, Cold atoms in dissipative optical lattices, Phys. Rep. {\bf 355}  335-451 (2001).


\bibitem{Chin}  C. Chin, R. Grimm, P. Julienne, and E. Tiesinga, Feshbach resonances in ultracold gases, Rev. Mod. Phys. {\bf 82}, 1225 (2010).

\bibitem{Cooper_review} N. R. Cooper, Rapidly rotating atomic gases, Adv. Phys. {\bf 57} 539-616 (2008).


\bibitem[{\citenamefont{Dalibard et~al.}(2011)\citenamefont{Dalibard, Gerbier,
  Juzeli{\=u}nas, and {\"O}hberg}}]{Dalibard2011}
\bibinfo{author}{\bibfnamefont{J.}~\bibnamefont{Dalibard}},
  \bibinfo{author}{\bibfnamefont{F.}~\bibnamefont{Gerbier}},
  \bibinfo{author}{\bibfnamefont{G.}~\bibnamefont{Juzeli{\=u}nas}},
  \bibnamefont{and}
  \bibinfo{author}{\bibfnamefont{P.}~\bibnamefont{{\"O}hberg}}, Colloquium:~Artificial gauge potentials for neutral atoms,
  \bibinfo{journal}{Rev. Mod. Phys.} \textbf{\bibinfo{volume}{83}},
  \bibinfo{pages}{1523} (\bibinfo{year}{2011}).

\bibitem[{\citenamefont{Goldman et~al.}(2014)\citenamefont{Goldman,
  Juzeli{\=u}nas, {\"O}hberg, and Spielman}}]{Goldman:2014bv}
\bibinfo{author}{\bibfnamefont{N.}~\bibnamefont{Goldman}},
  \bibinfo{author}{\bibfnamefont{G.}~\bibnamefont{Juzeli{\=u}nas}},
  \bibinfo{author}{\bibfnamefont{P.}~\bibnamefont{{\"O}hberg}},
  \bibnamefont{and} \bibinfo{author}{\bibfnamefont{I.~B.}
  \bibnamefont{Spielman}}, Light-induced gauge fields for ultracold atoms, \bibinfo{journal}{Rep. Prog. Phys.}
  \textbf{\bibinfo{volume}{77}}, \bibinfo{pages}{126401}
  (\bibinfo{year}{2014}).



\bibitem{GreinerMicroscope}
W.\ S.\ Bakr, J.\ I.\ Gillen, A.\ Peng, S.\ F\"olling, and M. Greiner, A quantum gas microscope for detecting single atoms in a Hubbard-regime optical lattice, 
Nature {\bf{462}}, 74-77 (2009).


\bibitem{Endres} M. Endres, M. Cheneau, T. Fukuhara, C. Weitenberg, P. Schau§, C. Gross, L. Mazza, M. C. Banuls, L.Pollet, I. Bloch and S. Kuhr, Single-site- and single-atom-resolved measurement of correlation functions, Appl. Phys. B {\bf 113} 27-39 (2013).


\bibitem{GreinerEntanglement}
R.\ Islam, R.\ Ma, P.\ M.\ Preiss, M.\ E.\ Tai, A.\ Lukin, M.\ Rispoli, and M.\ Greiner, Measuring entanglement entropy in a quantum many-body system, Nature {\bf 528}, 77 - 83 (2015).

\bibitem{Stamper1999} D. M. Stamper-Kurn, A. P. Chikkatur, A. Gšrlitz, S. Inouye, S. Gupta, D. E. Pritchard, and W. Ketterle, Excitation of Phonons in a Bose-Einstein Condensate by Light Scattering, Phys. Rev. Lett. {\bf 83} 2876-2879 (1999).

\bibitem[{\citenamefont{Jotzu et~al.}(2014)\citenamefont{Jotzu, Messer,
  Desbuquois, Lebrat, Uehlinger, Greif, and Esslinger}}]{jotzu2014}
\bibinfo{author}{\bibfnamefont{G.}~\bibnamefont{Jotzu}},
  \bibinfo{author}{\bibfnamefont{M.}~\bibnamefont{Messer}},
  \bibinfo{author}{\bibfnamefont{R.}~\bibnamefont{Desbuquois}},
  \bibinfo{author}{\bibfnamefont{M.}~\bibnamefont{Lebrat}},
  \bibinfo{author}{\bibfnamefont{T.}~\bibnamefont{Uehlinger}},
  \bibinfo{author}{\bibfnamefont{D.}~\bibnamefont{Greif}}, \bibnamefont{and}
  \bibinfo{author}{\bibfnamefont{T.}~\bibnamefont{Esslinger}}, Experimental realization of the topological Haldane model with ultracold fermions, 
  \bibinfo{journal}{Nature} \textbf{\bibinfo{volume}{515}},
  \bibinfo{pages}{237-240} (\bibinfo{year}{2014}).

\bibitem[{\citenamefont{Aidelsburger et~al.}(2015)\citenamefont{Aidelsburger,
  Lohse, Schweizer, Atala, Barreiro, Nascimb{\`e}ne, Cooper, Bloch, and
  Goldman}}]{Aidelsburger:2015}
\bibinfo{author}{\bibfnamefont{M.}~\bibnamefont{Aidelsburger}},
  \bibinfo{author}{\bibfnamefont{M.}~\bibnamefont{Lohse}},
  \bibinfo{author}{\bibfnamefont{C.}~\bibnamefont{Schweizer}},
  \bibinfo{author}{\bibfnamefont{M.}~\bibnamefont{Atala}},
  \bibinfo{author}{\bibfnamefont{J.~T.} \bibnamefont{Barreiro}},
  \bibinfo{author}{\bibfnamefont{S.}~\bibnamefont{Nascimb{\`e}ne}},
  \bibinfo{author}{\bibfnamefont{N.~R.} \bibnamefont{Cooper}},
  \bibinfo{author}{\bibfnamefont{I.}~\bibnamefont{Bloch}}, \bibnamefont{and}
  \bibinfo{author}{\bibfnamefont{N.}~\bibnamefont{Goldman}}, Measuring the Chern number of Hofstadter bands with ultracold bosonic atoms,
  \bibinfo{journal}{Nature Physics} \textbf{\bibinfo{volume}{11}},
  \bibinfo{pages}{162-166} (\bibinfo{year}{2015}).


\bibitem{Sorenson} A. S. S\o rensen, E. Demler, and M. D. Lukin, Fractional Quantum Hall States of Atoms in Optical Lattices, Phys. Rev. Lett. {\bf 94} 086803   (2005).

    \bibitem{Eckardt2}  A. Eckardt, T. Jinasundera, C. Weiss, and M. Holthaus, Analog of Photon-Assisted Tunneling in a Bose-Einstein Condensate, Phys. Rev. Lett. {\bf 95}, 200401 (2005). 

  \bibitem{Kitagawa} T. Kitagawa, E. Berg, M. Rudner, and E. Demler, Topological characterization of periodically driven quantum systems, Phys. Rev. B {\bf 82} 235114   (2010).

 \bibitem{Lindner} N. H. Lindner, G. Refael	 and V. Galitski, Floquet topological insulator in semiconductor quantum wells, Nature Phys. {\bf 7}  490-495   (2011).
 
   \bibitem{Bermudez} A. Bermudez, T. Schaetz, and D. Porras, Synthetic Gauge Fields for Vibrational Excitations of Trapped Ions, Phys. Rev. Lett. {\bf 107},
150501 (2011).

 
  \bibitem{Kolovsky} A. R. Kolovsky, Creating artificial magnetic fields for cold atoms by photon-assisted tunneling, Europhys. Lett. {\bf 93}  20003 (2011).

 
  \bibitem{Hauke2012} P. Hauke, O. Tieleman, A. Celi, C. \"Olschl\"ager, J. Simonet, J. Struck, M. Weinberg, P. Windpassinger, K. Sengstock, M. Lewenstein, and A. Eckardt, Non-Abelian Gauge Fields and Topological Insulators in Shaken Optical Lattices, Phys. Rev. Lett. {\bf 109}, 145301  (2012).

 \bibitem{Cayssol} J. Cayssol, B. D\'ora, F. Simon and R. Moessner, Floquet topological insulators, Phys. Status Solidi RRL {\bf 7}, 101-108 (2013). 

 \bibitem{GoldmanDalibard} N. Goldman and J. Dalibard, Periodically Driven Quantum Systems: Effective Hamiltonians and Engineered Gauge Fields, Phys. Rev. X {\bf 4}, 031027  (2014). 
 
  \bibitem{Zhai} W. Zheng and H. Zhai, Floquet topological states in shaking optical lattices, Phys. Rev. A {\bf 89}, 061603(R) (2014). 
 

 
 \bibitem{Bukov} M. Bukov, L. D'Alessio and A. Polkovnikov, Universal high-frequency behavior of periodically driven systems: from dynamical stabilization to Floquet engineering, Adv. Phys.,{\bf 64}, 139-226 (2015).

  \bibitem{Weitenberg} C. Weitenberg et al., Single-spin addressing in an atomic Mott insulator, Nature {\bf 471}, 319--324 (2011).
 
 
  


\bibitem{Juz:2004} G. Juzeli\=unas and P. \"Ohberg, Slow Light in Degenerate Fermi Gases, Phys. Rev. Lett. {\bf 93}, 033602 (2004).

\bibitem{Jaksch} D. Jaksch and P. Zoller, Creation of effective magnetic fields in optical lattices: the Hofstadter butterfly for cold neutral atoms, New J. Phys. {\bf 5} 56.1 - 56.11 (2003).

\bibitem{Osterloh:2005} K. Osterloh, M. Baig, L. Santos, P. Zoller, and M. Lewenstein, Cold Atoms in Non-Abelian Gauge Potentials: From the Hofstadter ``Moth" to Lattice Gauge Theory, Phys. Rev. Lett. {\bf 95}, 010403 (2005).

\bibitem{Juz:2005} J. Ruseckas, G. Juzeli\=unas, P. \"Ohberg, and M. Fleischhauer, Non-Abelian Gauge Potentials for Ultracold Atoms with Degenerate Dark States, Phys. Rev. Lett. {\bf 95}, 010404 (2005).




\bibitem[{\citenamefont{Celi et~al.}(2014)\citenamefont{Celi, Massignan,
  Ruseckas, Goldman, Spielman, Juzeli{\=u}nas, and Lewenstein}}]{Celi:2014}
\bibinfo{author}{\bibfnamefont{A.}~\bibnamefont{Celi}},
  \bibinfo{author}{\bibfnamefont{P.}~\bibnamefont{Massignan}},
  \bibinfo{author}{\bibfnamefont{J.}~\bibnamefont{Ruseckas}},
  \bibinfo{author}{\bibfnamefont{N.}~\bibnamefont{Goldman}},
  \bibinfo{author}{\bibfnamefont{I.~B.} \bibnamefont{Spielman}},
  \bibinfo{author}{\bibfnamefont{G.}~\bibnamefont{Juzeli{\=u}nas}},
  \bibnamefont{and}
  \bibinfo{author}{\bibfnamefont{M.}~\bibnamefont{Lewenstein}}, Synthetic Gauge Fields in Synthetic Dimensions, 
  \bibinfo{journal}{Phys. Rev. Lett.} \textbf{\bibinfo{volume}{112}},
  \bibinfo{pages}{043001} (\bibinfo{year}{2014}).

  \bibitem{Berryoriginal} M. V. Berry, Quantal Phase Factors Accompanying Adiabatic Changes, Proc. R. Soc. Lond. A {\bf 392}, 45-57 (1984).
  \bibitem{KarplusLuttinger}  R. Karplus and J. M. Luttinger, Hall Effect in Ferromagnetics, Phys. Rev. {\bf 95}, 1154-1160  (1954).
  \bibitem{MeadReview}  C. A. Mead, The geometric phase in molecular systems, Rev. Mod. Phys. {\bf 64} 51-85 (1992).

 \bibitem[{\citenamefont{Xiao et~al.}(2010)\citenamefont{Xiao, Chang, and
  Q.Niu}}]{Xiao2010}
\bibinfo{author}{\bibfnamefont{D.}~\bibnamefont{Xiao}},
  \bibinfo{author}{\bibfnamefont{M.-C.} \bibnamefont{Chang}}, \bibnamefont{and}
  \bibinfo{author}{\bibnamefont{Q. Niu}}, Berry phase effects on electronic properties,  \bibinfo{journal}{Rev. Mod. Phys}
  \textbf{\bibinfo{volume}{82}}, \bibinfo{pages}{1959} (\bibinfo{year}{2010}).

 \bibitem{PriceCooper} H. M. Price and N. R. Cooper, Mapping the Berry curvature from semiclassical dynamics in optical lattices, Phys. Rev. A {\bf 85}, 033620 (2012).

\bibitem{Duca} L. Duca, T. Li, M. Reitter, I. Bloch, M. Schleier-Smith, U. Schneider, An Aharonov-Bohm interferometer for determining Bloch band topology, Science {\bf 347}, 288-292 (2015).

  \bibitem{Alba} E. Alba, X. Fernandez-Gonzalvo, J. Mur-Petit, J. K. Pachos, and J. J. Garcia-Ripoll, Seeing Topological Order in Time-of-Flight Measurements, Phys. Rev. Lett. {\bf 107}, 235301 (2011).

 \bibitem{HaukeChern} P. Hauke, M. Lewenstein, and A. Eckardt, Tomography of Band Insulators from Quench Dynamics, Phys. Rev. Lett. {\bf 113}, 045303  (2014).

\bibitem{Flaschner} N. Fl\"aschner, B. S. Rem, M. Tarnowski, D. Vogel, D.-S. L\"uhmann, K. Sengstock, and C. Weitenberg, 	Experimental reconstruction of the Berry curvature in a Floquet Bloch band, Science {\bf 352},1091--1094 (2016).

  \bibitem[{\citenamefont{Nakahara}(2003)}]{Nakahara}
\bibinfo{author}{\bibfnamefont{M.}~\bibnamefont{Nakahara}},
  \emph{\bibinfo{title}{Geometry, Topology and Physics}}
  (\bibinfo{publisher}{IOP Publishing Ltd.}, \bibinfo{address}{Bristol and
  Philadelphia}, \bibinfo{year}{2003}).



\bibitem{TKNN1982}
\bibinfo{author}{\bibfnamefont{D.~J.} \bibnamefont{Thouless}},
  \bibinfo{author}{\bibfnamefont{M.}~\bibnamefont{Kohmoto}},
  \bibinfo{author}{\bibfnamefont{M.~P.} \bibnamefont{Nightingale}},
  \bibnamefont{and} \bibinfo{author}{\bibfnamefont{M.}~\bibnamefont{den Nijs}}, Quantized Hall Conductance in a Two-Dimensional Periodic Potential, 
  \bibinfo{journal}{Phys. Rev. Lett.} \textbf{\bibinfo{volume}{49}},
  \bibinfo{pages}{405-408} (\bibinfo{year}{1982}).

\bibitem{Dauphin:2013} A. Dauphin and N. Goldman, Extracting the Chern Number from the Dynamics of a Fermi Gas: Implementing a Quantum Hall Bar for Cold Atoms, Phys. Rev. Lett. {\bf 111}, 135302 (2013).

\bibitem{Price:2016} H. M. Price, O. Zilberberg, T. Ozawa, I. Carusotto, and N. Goldman, Measurement of Chern numbers through center-of-mass responses, Phys. Rev. B {\bf 93}, 245113 (2016).


  \bibitem{Liu:2013} X.-J. Liu, K. T. Law, T. K. Ng, and P. A. Lee, Detecting Topological Phases in Cold Atoms, Phys. Rev. Lett. {\bf 111} 120402 (2013).
  
\bibitem{2DSOC2:2015} Z. Wu, L. Zhang, W. Sun, X.-T. Xu, B.-Z. Wang, S.-C. Ji, Y. Deng, S. Chen, X.-J. Liu, J.-W. Pan, Realization of Two-Dimensional Spin-orbit Coupling for Bose-Einstein Condensates, arXiv:1511.08170 (2015).
  
  
\bibitem{Chernimpurity}  F.  Grusdt, N. Y.  Yao, D. Abanin, M.  Fleischhauer,and  E.  Demler, Interferometric measurements of many-body topological invariants using mobile impurities, Nature Comm. {\bf 7}, 11994 (2016).


\bibitem{Halperin1982}
B. I. Halperin, Quantized Hall conductance, current-carrying edge states, and the existence of extended states in a two-dimensional disordered potential, Phys. Rev. B {\bf{25}}, 2185-2190 (1982).



\bibitem{GoldmanPNAS} N. Goldman, J. Dalibard, A. Dauphin, F. Gerbier, M. Lewenstein, P. Zoller, and I. B. Spielman, Direct imaging of topological edge states in cold-atom systems, PNAS {\bf 110}, 6736-6741 (2013).

\bibitem{MuellerEdge} M. D. Reichl and E. J. Mueller, Floquet edge states with ultracold atoms, Phys. Rev. A {\bf 89}, 063628 (2014).

\bibitem{Mancini:2015} M. Mancini, G. Pagano, G. Cappellini, L. Livi, M. Rider, J. Catani, C. Sias, P. Zoller, M. Inguscio, M. Dalmonte, and L. Fallani, Observation of chiral edge states with neutral fermions in synthetic Hall ribbons, Science {\bf 349}, 1510-1513 (2015).

\bibitem{Stuhl:2015} B. K. Stuhl, H. I. Lu, L. M. Aycock, D. Genkina, and I. B. Spielman, Visualizing edge states with an atomic Bose gas in the quantum Hall regime, Science {\bf 349},  1514-1518 (2015).



\bibitem[{\citenamefont{Atala et~al.}(2014)\citenamefont{Atala, Aidelsburger,
  Lohse, Barreiro, Paredes, and Bloch}}]{Atala:2014}
\bibinfo{author}{\bibfnamefont{M.}~\bibnamefont{Atala}},
  \bibinfo{author}{\bibfnamefont{M.}~\bibnamefont{Aidelsburger}},
  \bibinfo{author}{\bibfnamefont{M.}~\bibnamefont{Lohse}},
  \bibinfo{author}{\bibfnamefont{J.~T.} \bibnamefont{Barreiro}},
  \bibinfo{author}{\bibfnamefont{B.}~\bibnamefont{Paredes}}, \bibnamefont{and}
  \bibinfo{author}{\bibfnamefont{I.}~\bibnamefont{Bloch}}, Observation of chiral currents with ultracold atoms in bosonic ladders, 
  \bibinfo{journal}{Nature Phys.} \textbf{\bibinfo{volume}{10}},
  \bibinfo{pages}{588-593} (\bibinfo{year}{2014}).

\bibitem{Liuspectro} X.-J. Liu, X. Liu, C. Wu, and J. Sinova, Quantum anomalous Hall effect with cold atoms trapped in a square lattice, Phys. Rev. A {\bf 81}, 033622 (2010).

\bibitem{Stanescu} T. D. Stanescu, V. Galitski, and S. Das Sarma, Topological states in two-dimensional optical lattices, Phys. Rev. A {\bf 82}, 013608 (2010).

\bibitem{Goldmanspectro} N. Goldman, J. Beugnon, and F. Gerbier, Detecting Chiral Edge States in the Hofstadter Optical Lattice, Phys. Rev. Lett. {\bf 108}, 255303 (2012).


\bibitem{Bermudez2010} A. Bermudez, L. Mazza, M. Rizzi, N. Goldman, M. Lewenstein and
M.A. Martin-Delgado, Wilson Fermions and Axion Electrodynamics in Optical Lattices, Phys. Rev. Lett. {\bf 105}, 190404 (2010) .

\bibitem{WeylAtoms} T. Dubcek, C. J. Kennedy, L. Lu, W. Ketterle, M. Soljacic, and H. Buljan, Weyl Points in Three-Dimensional Optical Lattices: Synthetic Magnetic Monopoles in Momentum Space, 
Phys. Rev. Lett. {\bf 114}, 225301 (2015).



\bibitem{4Datoms:2015} H. M. Price, O. Zilberberg, T. Ozawa, I. Carusotto and N. Goldman,  Four-Dimensional Quantum Hall Effect with Ultracold Atoms, Phys. Rev. Lett. {\bf 115}, 195303 (2015).




\bibitem{Luttinger} J. M. Luttinger, The Effect of a Magnetic Field on Electrons in a Periodic Potential, Phys. Rev. {\bf 84}, 814--817  (1951).


\bibitem[{\citenamefont{Hofstadter}(1976)}]{Hofstadter}
\bibinfo{author}{\bibfnamefont{D.~R.} \bibnamefont{Hofstadter}},
 \bibinfo{titel}{Energy levels and wave functions of Bloch electrons in rational and irrational magnetic fields},
  \bibinfo{journal}{Phys. Rev. B} \textbf{\bibinfo{volume}{14}},
  \bibinfo{pages}{2239--2249} (\bibinfo{year}{1976}).


\bibitem{CooperFluxLattice}
N.\ R.\ Cooper, Optical Flux Lattices for Ultracold Atomic Gases,
Phys.\ Rev.\ Lett.\ {\bf{106}}, 175301 (2011).


\bibitem{Jaksch1998} D. Jaksch, C. Bruder, J. I. Cirac,  C. W. Gardiner,  and P. Zoller, Cold Bosonic Atoms in Optical Lattices, Phys. Rev. Lett. {\bf 81} 3108--3111 (1998).

\bibitem{Ruostekoski} J. Ruostekoski, G. V. Dunne, and J. Javanainen, Manipulating atoms in an optical lattice: Fractional fermion number and its optical quantum measurement,  Phys. Rev.
Lett. {\bf 88}, 180401 (2002).
  
\bibitem{Gerbier} F. Gerbier and J. Dalibard, Gauge fields for ultracold atoms in
optical superlattices, New J. Phys. {\bf 12}, 033007 (2010).
  
  
\bibitem{Goldman2010} N. Goldman, I. Satija, P. Nikolic, A. Bermudez, M.A. Martin-Delgado,
M. Lewenstein, and I. B. Spielman, Realistic Time-Reversal Invariant Topological Insulators with Neutral Atoms, Phys. Rev. Lett. {\bf 105}, 255302 (2010).



\bibitem{Liu2014} X.-J. Liu, K. T. Law, and T. K. Ng, Realization of 2D Spin-Orbit Interaction and Exotic Topological Orders in Cold Atoms, Phys. Rev. Lett. {\bf 112}, 086401 (2014).



\bibitem{YaoFlatDipolar}
N.\ Y.\ Yao, C.\ R.\ Laumann, A.\ V.\ Gorshkov, S.\ D.\ Bennett, E.\ Demler, P.\ Zoller, and M.\ D.\ Lukin, Topological Flat Bands from Dipolar Spin Systems, Phys.\ Rev.\ Lett.\ {\bf{109}}, 266804 (2012).

\bibitem{Sun2011} K. Sun, W. V. Liu, A. Hemmerich, S. Das Sarma, Topological semimetal in a fermionic optical lattice, Nature Physics {\bf 8}, 67--70 (2012).

 \bibitem{DauphinMott} A. Dauphin, M. M\"uller, M. A. Martin-Delgado, Rydberg-atom quantum simulation and Chern-number characterization of a topological Mott insulator, Phys. Rev. A {\bf 86}, 053618 (2012).
 

 \bibitem{Barbarino} S. Barbarino, L. Taddia, D. Rossini, L. Mazza, R. Fazio, Magnetic crystals and helical liquids in alkaline-earth fermionic gases, Nat. Commun. {\bf 6}, 8134 (2015).

\bibitem{Lacki} M.  Lacki, H. Pichler, A. Sterdyniak, A.s Lyras, V. E. Lembessis, O. Al-Dossary, J. C. Budich, P. Zoller, Quantum Hall physics with cold atoms in cylindrical optical lattices, Phys. Rev. A {\bf 93}, 013604 (2016).
 
 



\bibitem[{\citenamefont{Aidelsburger et~al.}(2013)\citenamefont{Aidelsburger,
  Atala, Lohse, Barreiro, Paredes, and Bloch}}]{aidelsburger2013}
\bibinfo{author}{\bibfnamefont{M.}~\bibnamefont{Aidelsburger}},
  \bibinfo{author}{\bibfnamefont{M.}~\bibnamefont{Atala}},
  \bibinfo{author}{\bibfnamefont{M.}~\bibnamefont{Lohse}},
  \bibinfo{author}{\bibfnamefont{J.~T.} \bibnamefont{Barreiro}},
  \bibinfo{author}{\bibfnamefont{B.}~\bibnamefont{Paredes}}, \bibnamefont{and}
  \bibinfo{author}{\bibfnamefont{I.}~\bibnamefont{Bloch}},
  \bibinfo{titel}{Realization of the Hofstadter Hamiltonian with Ultracold Atoms in Optical Lattices},
  \bibinfo{journal}{Phys. Rev. Lett.} \textbf{\bibinfo{volume}{111}},
  \bibinfo{pages}{185301} (\bibinfo{year}{2013}).

\bibitem[{\citenamefont{Miyake et~al.}(2013)\citenamefont{Miyake, Siviloglou,
  Kennedy, Burton, and Ketterle}}]{miyake2013}
\bibinfo{author}{\bibfnamefont{H.}~\bibnamefont{Miyake}},
  \bibinfo{author}{\bibfnamefont{G.~A.} \bibnamefont{Siviloglou}},
  \bibinfo{author}{\bibfnamefont{C.~J.} \bibnamefont{Kennedy}},
  \bibinfo{author}{\bibfnamefont{W.~C.} \bibnamefont{Burton}},
  \bibnamefont{and} \bibinfo{author}{\bibfnamefont{W.}~\bibnamefont{Ketterle}},
  \bibinfo{titel}{Realizing the Harper Hamiltonian with Laser-Assisted Tunneling in Optical Lattices},
  \bibinfo{journal}{Phys. Rev. Lett.} \textbf{\bibinfo{volume}{111}}, 
  \bibinfo{pages}{185302} (\bibinfo{year}{2013}).

 \bibitem{Kennedy} C. J. Kennedy,	W. C. Burton,	W. C. Chung and W. Ketterle, Observation of BoseÐEinstein condensation in a strong synthetic magnetic field, Nature Phys. {\bf 11},  859-864 (2015).



 \bibitem{Lohse} M. Lohse, C. Schweizer, O. Zilberberg, M. Aidelsburger and I. Bloch, A Thouless quantum pump with ultracold bosonic atoms in an optical superlattice, Nature Phys. {\bf{12}}, 350--354 (2016).

\bibitem{Nakajima} S. Nakajima, T. Tomita, S. Taie, T. Ichinose, H. Ozawa, L. Wang, M. Troyer, Y. Takahashi, Topological Thouless pumping of ultracold fermions, Nature Phys. {\bf{12}}, 296--300 (2015).

 \bibitem{Lu2015} H.-I. Lu, M. Schemmer, L. M. Aycock, D. Genkina, S. Sugawa, and I. B. Spielman, Geometrical Pumping with a Bose-Einstein Condensate, Phys. Rev. Lett. {\bf 116}, 200402 (2016).
 
\bibitem{Thouless1983} 
D. J. Thouless, Quantization of particle transport,
Phys. Rev. B {\bf{27}},  6083--6087 (1983). 

 \bibitem{Struck} J. Struck, M. Weinberg, C. …lschlŠger, P. Windpassinger, J. Simonet,	K. Sengstock, R. H\"oppner, P. Hauke,	A. Eckardt, M. Lewenstein and L. Mathey, Engineering Ising-XY spin-models in a triangular lattice using tunable artificial gauge fields, Nature Phys. {\bf 9}, 738--743 (2013). 


  \bibitem{Haldane} F. D. M. Haldane, Model for a Quantum Hall Effect without Landau Levels: Condensed-Matter Realization of the "Parity Anomaly", Phys. Rev. Lett. {\bf 61},  2015-2018 (1988).









\bibitem{MooreRead}
G. Moore and N. Read, Nonabelions In The Fractional Quantum Hall Effect,  Nucl. Phys. B {\bf{360}}, 362--396 (1991).

\bibitem{ReadGreen}
N. Read and D. Green, Paired states of fermions in two dimensions with breaking of parity and time-reversal symmetries and the fractional quantum Hall effect, Phys. Rev. B {\bf{61}}, 10267--10297 (2000).

\bibitem{Kitaev2001}
	A.\ Kitaev, Unpaired Majorana fermions in quantum wires,
  	Phys.-Usp. {\bf 44}, 131--136 (2001).


\bibitem{SauRashbaZeeman}
J.\ D.\ Sau, R.\ M.\ Lutchyn, S.\ Tewari, and S.\ D.\ Sarma, Generic New Platform for Topological Quantum Computation Using Semiconductor Heterostructures, Phys.\ Rev.\ Lett.\ {\bf{104}}, 040502 (2010).


\bibitem{SpielmanReview}
V.\ Galitski, I.\ B.\ Spielman, SpinÐorbit coupling in quantum gases, Nature {\bf{494}}, 49--54 (2013).

\bibitem{2DSOCNatPhys}
 L. Huang, Z. Meng, P. Wang, P. Peng, S.-L. Zhang, L. Chen, D. Li, Q. Zhou, and J. Zhang,
 Experimental realization of two-dimensional synthetic spin-orbit coupling in ultracold Fermi gases, Nature Physics {\bf 12}, 540--544 (2016).


\bibitem{pwaveDasSarma}
C.\ Zhang, S.\ Tewari, R.\ M.\ Lutchyn, S.\ Das\ Sarma, $p_x+ip_y$ Superfluid from $s$-Wave Interactions of Fermionic Cold Atoms, Phys. Rev. Lett. {\bf{101}}, 160401 (2008).

\bibitem{pwaveLewenstein}
P.\ Massignan, A.\ Sanpera, M.\ Lewenstein, Creating $p$-wave superfluids and topological excitations in optical lattices, Phys. Rev. A {\bf{81}}, 031607(R) (2010).

\bibitem{pwaveMelo}
K.\ Seo, L.\ Han, and C.\ S\'{a} de Melo, Emergence of Majorana and Dirac Particles in Ultracold Fermions via Tunable Interactions, Spin-Orbit Effects, and Zeeman Fields, Phys. Rev. Lett. {\bf{109}}, 105303 (2012).




\bibitem{Tewaripip}
S.\ Tewari, S.\ Das Sarma, C.\ Nayak, C.\ Zhang, and P.\ Zoller, 
Quantum Computation using Vortices and Majorana Zero Modes of a $p_x+i p_y$ Superfluid of Fermionic Cold Atoms, Phys. Rev. Lett. {\bf 98}, 010506 (2007).

\bibitem{Jiang11}
L.\ Jiang, T.\ Kitagawa, J.\ Alicea, A.\ R.\ Akhmerov, D.\ Pekker, G.\ Refael, J.\ I.\ Cirac, E.\ Demler, M.\ D.\ Lukin, and P.\ Zoller, Majorana Fermions in Equilibrium and in Driven Cold-Atom Quantum Wires,
Phys.\ Rev.\ Lett.\ {\bf{106}}, 220402 (2011).

  
 \bibitem{Nascimbene12}
\bibinfo{author}{\bibfnamefont{S.}~\bibnamefont{Nascimb\`{e}ne}},
\bibinfo{titel}{Realizing one-dimensional topological superfluids with ultracold atomic gases},
  \bibinfo{journal}{Journal of Physics B} \textbf{\bibinfo{volume}{46}},
  \bibinfo{pages}{134005} (\bibinfo{year}{2013}).   
  
\bibitem{Kraus13}
\bibinfo{author}{\bibfnamefont{C.}~\bibfnamefont{V.}~\bibnamefont{Kraus}},
\bibinfo{author}{\bibfnamefont{M.}~\bibnamefont{Dalmonte}},
\bibinfo{author}{\bibfnamefont{M.}~\bibfnamefont{A.}~\bibnamefont{Baranov}},
\bibinfo{author}{\bibfnamefont{A.}~\bibfnamefont{M.}~\bibnamefont{L\"auchli}},
  \bibnamefont{and} \bibinfo{author}{\bibfnamefont{P.}~\bibnamefont{Zoller}},
\bibinfo{titel}{Majorana Edge States in Atomic Wires Coupled by Pair Hopping},
  \bibinfo{journal}{Phys. Rev. Lett.} \textbf{\bibinfo{volume}{111}},
  \bibinfo{pages}{173004} (\bibinfo{year}{2013}).   

\bibitem{KrausBraiding}  
C.\ V.\ Kraus, P.\ Zoller, and M.\ A.\ Baranov, Braiding of Atomic Majorana Fermions in Wire Networks and Implementation of the Deutsch-Jozsa Algorithm, Phys.\ Rev.\ Lett.\ {\bf{111}}, 203001 (2013).


\bibitem{DiehlDissPrep}
S.~Diehl, A.~Micheli, A.~Kantian, B.~Kraus, H.~ P.~ B\"uchler, and P.~Zoller, Quantum states and phases in driven open quantum systems with cold atoms,
Nature Physics {\bf{4}}, 878--883 (2008).

\bibitem{Verstraete2008}
F.\ Verstraete, M.\ M.\ Wolf, J.\ I.\ Cirac, Quantum computation and quantum-state engineering driven by dissipation
Nature Physics {\bf{5}}, 633--636 (2009).


\bibitem{Lindblad1976}
G. Lindblad, On the generators of quantum dynamical semigroups, Commun. Math. Phys. {\bf 48} (2), 119--130 (1976).

\bibitem{DiehlTopDiss}
S.~Diehl, E.~Rico, M.~A. Baranov, and P.~Zoller, Topology by dissipation in atomic quantum wires,
Nature Physics {\bf{7}}, 971--977 (2011).




\bibitem{BardynTopDiss} 
	C.-E.\ Bardyn, M.\ A.\ Baranov, C.\ V.\ Kraus, E.\ Rico, A.\ Imamoglu, P.\ Zoller, S.\ Diehl, Topology by dissipation, New J. Phys. {\bf 15}, 085001  (2013).



\bibitem{Bardyn2012}
C.-E.\ Bardyn, M.\ A.\ Baranov, E.\ Rico, A.\ Imamoglu, P.\ Zoller, S.\ Diehl, Majorana Modes in Driven-Dissipative Atomic Superfluids with a Zero Chern Number,
Phys.\ Rev.\ Lett.\ {\bf{109}}, 130402 (2012). 

\bibitem{chernalgebraic} 
	D.\ J.\ Thouless, 
Wannier functions for magnetic sub-bands, J.\ Phys.\ C {\bf 17},  L325--L327 (1984).

%
	
\bibitem{DissCI}
J.\ C.\ Budich, P.\ Zoller, S.\ Diehl, Dissipative preparation of Chern insulators, Phys. Rev. A {\bf{91}},  042117 (2015).

\bibitem{TopDens}
J.\ C.\ Budich, S.\ Diehl, Topology of density matrices, Phys. Rev. B {\bf{91}}, 165140 (2015).	




 
\bibitem{Wen2007}
X.-G. Wen, {\it Quantum Field Theory of Many-body Systems} (Oxford University Press, 2007). 

\bibitem{HannesEntanglementGrowth}  
A.\ J.\ Daley, H.\ Pichler, J.\ Schachenmayer, and P.\ Zoller, Measuring Entanglement Growth in Quench Dynamics of Bosons in an Optical Lattice, Phys.\ Rev.\ Lett.\ {\bf{109}}, 020505 (2015).




 
\bibitem{PalmerJaksch}
R.\ N.\ Palmer, D.\ Jaksch, High-Field Fractional Quantum Hall Effect in Optical Lattices, Phys. Rev. Lett. {\bf{96}}, 180407 (2006). 
 
\bibitem{SorensenHafezi}
M.\ Hafezi, A.\ S.\ S\o rensen, E.\ Demler, M.\ D.\ Lukin, Fractional quantum Hall effect in optical lattices, Phys. Rev. A {\bf{76}}, 023613 (2007).



\bibitem{MollerCooper}
G.\ Moller, N.\ R.\ Cooper, Composite Fermion Theory for Bosonic Quantum Hall States on Lattices, Phys. Rev. Lett. {\bf{103}}, 105303 (2009). 

\bibitem{KapitMueller}
E.\ Kapit, E.\ Mueller, Exact Parent Hamiltonian for the Quantum Hall States in a Lattice, Phys. Rev. Lett. {\bf{105}}, 215303 (2010). 

\bibitem{NielsenCirac}
A.\ E.\ B.\ Nielsen, G.\ Sierra, J.\ I.\ Cirac, Local models of fractional quantum Hall states in lattices and physical implementation, Nat. Commun. {\bf{4}}, 2864 (2013). 
 


\bibitem{CooperFluxFlat}
N.\ R.\ Cooper and J.\ Dalibard, Reaching Fractional Quantum Hall States with Optical Flux Lattices,
Phys.\ Rev.\ Lett.\ {\bf{110}}, 185301 (2013).



\bibitem{MollerHigherChern}
G.\ M\"oller and N.\ R.\ Cooper, Fractional Chern Insulators in Harper-Hofstadter Bands with Higher Chern Number, Phys.\ Rev.\ Lett.\ {\bf{115}}, 126401 (2015).
 
 \bibitem{Dai} H.-N. Dai, B. Yang, A. Reingruber, H. Sun, X.-F. Xu, Y.-A. Chen, Z.-S. Yuan, J.-W. Pan, Observation of Four-body Ring-exchange Interactions and Anyonic Fractional Statistics, arXiv:1602.05709 (2016).

\bibitem{Kitanew} A.Yu. Kitaev, Fault-tolerant quantum computation by anyons, Ann. Phys. {\bf 303} 2--30 (2003).
 
\bibitem{AlexFQM}  
A.\ W.\ Glaetzle, M.\ Dalmonte, R.\ Nath, C.\ Gross, I.\ Bloch, and P.\ Zoller, Designing Frustrated Quantum Magnets with Laser-Dressed Rydberg Atoms, Phys.\ Rev.\ Lett.\ {\bf{114}}, 173002 (2015).  

\bibitem{PohlFQM}
R.\ M.\ W.\ van Bijnen and T.\ Pohl, Phys. Rev. Lett. {\bf{114}}, Quantum Magnetism and Topological Ordering via Rydberg Dressing near F\"orster Resonances, 243002 (2015).

\bibitem{AlexQSI} 
A.\ W.\ Glaetzle, M.\ Dalmonte, R.\ Nath, I.\ Rousochatzakis, R.\ Moessner, and P.\ Zoller, Quantum Spin-Ice and Dimer Models with Rydberg Atoms, Phys. Rev. X {\bf{4}}, 041037 (2014).


\end{thebibliography}
\end{document}